\documentclass[letter, twocolumn]{jpsj2} 
\title{
Quasi-Classical Calculation of the Mixed-State 
Thermal Conductivity \\
in $s$-Wave and $d$-Wave Superconductors} 

\author{Hiroto \textsc{ADACHI}
\thanks{E-mail address: adachi@itp.phys.ethz.ch} 
\thanks{Present address: Theoretische Physik, ETH-H\"{o}nggerberg, 
CH-8093 Z\"{u}rich, Switzerland}, 
Predrag \textsc{Miranovi\'{c}}$^{1}$, 
Masanori \textsc{ICHIOKA}, 
and Kazushige \textsc{MACHIDA}}

\inst{
Department of Physics, Okayama University, Okayama 700-8530 \\
$^{1}$
Department of Physics, University of Montenegro, Podgorica 81000, 
Montenegro
}

\abst{
To see how superconducting gap structures affect 
the longitudinal component of mixed-state thermal conductivity $\kappa_{xx}(B)$, 
the magnetic-field dependences of 
$\kappa_{xx}(B)$ in $s$-wave and $d$-wave superconductors are investigated. 
Calculations are performed on the basis of the 
quasi-classical theory of superconductivity 
by fully taking account of the spatial variation of the normal Green's function, 
neglected in previous works, by the Brandt-Pesch-Tewordt approximation. 
On the basis of our result, we discuss the possibility of 
$\kappa_{xx}(B)$ measurement 
as a method of probing the gap structure. 
}

\kword{quasi-classical theory, thermal conductivity, mixed state, 
$s$-wave superconductor$, d$-wave superconductor} 

\begin{document}
\maketitle

\section{Introduction} 
In the last two decades, several new superconductors 
showing unconventional pairing properties 
have been found among heavy-fermion, 
oxide, and organic materials.~\cite{Sigrist} 
These discoveries have motivated 
studies on 
identifying the gap symmetry, 
as well as clarifying the mechanism of the superconductivity. 
At present, it has been well established that bulk measurements of 
mixed-state thermodynamic quantities can provide important 
information on the gap structure. 
~\cite{Caroli,Volovik,Won1,Kopnin,Simon,Vekhter1} 
The principle behind this technique is that 
the low-energy quasiparticles around vortex cores, 
relevant to the low-temperature thermodynamics, 
sensitively reflect the structure of the superconducting gap. 
On the basis of the numerical solution of quasi-classical Eilenberger equations, 
we have demonstrated that 
the gap structure can be inferred from the low-temperature field 
dependence~\cite{Ichioka,Nakai} 
and field-angle dependence~\cite{Pedja1a,Pedja1b,Adachia,Adachib} 
of the specific heat and magnetization. 

Thermal-transport measurement in a superconducting mixed state is another option 
for such a spectroscopic method based on the bulk property of a sample. 
At very low temperatures, where the transport 
property is mainly determined by the elastic impurity scattering, 
the thermal conductivity can also provide important information on the pairing state. 
Roughly speaking, the longitudinal component of mixed-state 
thermal conductivity $\kappa_{xx}(B)$ 
at zero temperature is given by 
$\kappa_{xx}(B) \sim \langle v_x^2 N_{\mib{p}}(0) 
\tau^{\rm tr}_{\mib{p}}(0) \rangle$, 
where $N_{\mib{p}}(0)$ is the momentum-resolved density of states at the Fermi surface, 
$v_x$ is the $x$-component of the Fermi velocity, $\tau^{\rm tr}_{\mib{ p}}(0)$ 
is the momentum-dependent transport lifetime, and $\langle \cdots \rangle$ 
denotes the average over the Fermi surface (see eq.~(\ref{Eq:Boltzmann}) below). 
As Volovik~\cite{Volovik} pointed out, 
$N_{\mib{p}}(0)$ and $\tau^{\rm tr}_{\mib{p}}(0)$ 
in a nodal superconductor are modified by the field-induced Doppler shift 
on the delocalized quasiparticles, while in an $s$-wave superconductor, 
such an effect is negligibly small. 
Hence, we expect that the behavior of $\kappa_{xx}(B)$ in a nodal superconductor 
is different from that in an $s$-wave superconductor. 
Recently, a series of mixed-state thermal-transport experiments have been extensively 
performed by Matsuda's group to determine the gap structures in 
several newly found superconductors.~\cite{Matsuda1} 

The low-temperature mixed-state thermal transport is mainly determined by 
the following three mechanisms: 
impurity scattering, giving the Drude thermal conductivity in 
the normal state; Andreev scattering by vortex cores;~\cite{Andreev} 
and the Doppler shift due to the supercurrents.~\cite{Maki1} 
So far, there have been several approaches to the calculation of the 
mixed-state thermal conductivity in a moderately clean superconductor.
~\cite{Maki2,Cleary,Houghton,Yu,Franz,Kubert,Barash,Won2,Vekhter,Takigawa,Dukan,Durst} 
Among them, two approximations have been frequently used. 
One is the Doppler shift approximation.~\cite{Kubert,Barash} 
This method neglects the Andreev scattering by vortex cores, and 
thus is valid only near the lower critical field $H_{c1}$. 
The other approximation~\cite{Houghton,Vekhter} 
is based on the Brandt-Pesch-Tewordt (BPT) approximation.~\cite{BPT,Pesch}
The latter is superior to the former in that 
it takes account of both the Doppler shift and 
the vortex-core scattering. 
In the BPT approximation, however, the spatial variation of 
the normal Green's function is neglected, 
which is strongly related to the local density of states 
detected by the scanning tunneling microscopy experiment.~\cite{Hess} 
Hence, this approximation is valid only near the upper critical field $H_{c2}$. 
At present, there is no theoretical work valid 
in the wide field range from $H_{c1}$ to $H_{c2}$. 
In view of analyzing the experimental data for several newly found superconductors, 
it is of importance to calculate the mixed-state thermal conductivity 
beyond these two approximations. 

It is well known that the quasi-classical theory of 
superconductivity~\cite{Eilenberger,LO1} 
can accurately describe the mixed-state properties in a wide field range 
from $H_{c1}$ to $H_{c2}$. 
The advantage of this framework is that, 
if applied to the calculation of mixed-state thermal conductivity, 
it can take account of the spatial variation of the normal Green's function 
neglected in the BPT approximation, 
as well as the Andreev vortex scattering neglected in the Doppler shift 
approximation. 
Of course, we need to solve a set of transport-like equations 
to accomplish the computation. 

In this work, we adopt the quasi-classical theory of superconductivity, and 
develop a method of calculating the mixed-state thermal conductivity $\kappa_{xx}(B)$. 
Then, we apply our analysis to two-dimensional $s$-wave and $d$-wave superconductors, 
and calculate the magnetic-field dependences of $\kappa_{xx}(B)$ 
to clarify the effect of the gap structure 
on the mixed-state thermal transport. 
We also study the effect of the in-plane Fermi surface anisotropy. 
On the basis of our result, we discuss the possibility of using $\kappa_{xx}(B)$ 
as a method of probing the gap structure. 

The paper is organized as follows. 
In \S 2, we develop a method of calculating the mixed-state 
thermal conductivity based on the quasi-classical theory of superconductivity. 
In \S 3, we present our numerical results for the mixed-state 
thermal conductivity obtained by solving the quasi-classical equations. 
We discuss our result in \S 4, 
and conclude in \S 5. 
We use the unit $\hbar=c=k_{\rm B}=1$ throughout this paper.

\section{Formulation}
\subsection{Linear-response-equation} 
Our approach to mixed-state thermal transport 
is based on the Kubo formula for a homogeneous temperature gradient.~\cite{Kubo} 
Then, the mixed-state thermal conductivity $\kappa_{xx}(B)$ 
is given by 
\begin{equation} 
  \kappa_{xx}(B) = -\frac{1}{T} {\rm Im} 
  \left.
  \frac{Q^R_{xx}(\omega) }{\omega} 
  \right|_{\omega \to 0} , \label{Eq:Kubo}
\end{equation} 
where 
$Q^R_{xx}$ 
is obtained from the (Matsubara) heat-current correlation function 
\begin{eqnarray}
  Q_{xx}({\rm i}\omega_m) &=& 
  \frac{1}{V} \int d^2 r_1 d^2 r_2 
  \int_0^{T^{-1}} d \tau 
  e^{{\rm i} \omega_m \tau} \nonumber \\
  && \times 
  \Big( -
  \big{\langle} T_\tau[J_x(\mib{r}_1,\tau)J_x(\mib{r}_2,0)] \big{\rangle}_{\rm eq} 
  \Big), 
\end{eqnarray}
after the analytical continuation 
$Q_{xx}^R(\omega) = Q_{xx}({\rm i}\omega_m \to \omega + {\rm i}0_{+})$. 
Here, $\omega_m= 2 \pi T m$ are bosonic Matsubara frequencies, 
$\langle \cdots \rangle_{\rm eq}$ denotes the statistical average, and 
$V$ is the volume of the sample. 
The heat-current operator $\mib{J}$ in the real-time representation is given by 
\begin{eqnarray}
  \mib{J}({\it 1}) &=& 
  -\frac{1}{2m} \sum_{\rm spin} \Big[ 
  \partial_{t_1} \big(\mib{\nabla}_{r_2}+ {\rm i}|e|\mib{A}({\it 2}) \big) \nonumber \\ 
  && 
  \left. \quad + 
  \big(\mib{\nabla}_{r_{1}}- {\rm i}|e|\mib{A}({\it 1}) \big) \partial_{t_{2}}
  \Big] \psi^\dag({\it 1}) \psi({\it 2}) \right|_{{\it 2} \to {\it 1}}, 
\end{eqnarray} 
where $\psi$ is the electron-field operator and 
${\it 1}=(\mib{r}_1, t_1), {\it 2}=(\mib{r}_2, t_2)$. 
To evaluate $Q_{xx}$, we follow the 
procedure suggested by Klimesch and Pesch.~\cite{Klimesch} 
They showed that the heat-current correlation function can be constructed in a 
manner similar to obtaining the density correlation function $D$ 
under charge disturbance. 
On the imaginary (Matsubara) axis, $D$ is given by 
\begin{eqnarray} 
  D({\rm i}\omega_m) &=& 
  \frac{1}{V} \int d^2 r_1 d^2 r_2 
  \int_0^{T^{-1}} d \tau 
  e^{{\rm i} \omega_m \tau} \nonumber \\
  && \times 
  \Big( -
    \big{\langle} T_\tau[n(\mib{r}_1,\tau)n(\mib{r}_2,0)] \big{\rangle}_{\rm eq} 
  \Big),  
\end{eqnarray}
where 
$n = \sum_{\rm spin} \psi^\dag \psi$ is the electron density. 
The procedure by Klimesch and Pesch is justified 
as long as we consider a moderately clean superconductor 
$\xi_0/l \ll 1$ (coherence length $\xi_0$ and mean free path $l$), 
where the impurity vertex corrections for the heat current can be safely neglected. 
When we consider a dirtier superconductor, we have to 
include the vertex corrections to satisfy the conservation of 
the energy density.~\cite{comm1,Ambegaokar,Hirschfeld} 
A full quasi-classical treatment for the charge current 
that includes the impurity vertex corrections 
can be found in ref.~\citen{Eschrig}. 
Note that the present result can be also derived 
from the Keldysh technique.~\cite{Pedja2} 

Let us briefly review the procedure used by Klimesch and Pesch~\cite{Klimesch} to 
find the connection between the heat-current correlation function $Q_{xx}$ 
and the density correlation function $D$. 
Consider an external scalar potential $\Phi= -|e|\phi$ 
that couples to the electron density ${n}$. 
Next, we define the quasi-classical Green's function as 
\begin{equation}
  \widehat{g} = 
  \int \frac{d \zeta_{\mib{p}} }{{\rm i}\pi}  
  \left( { G \atop  -F^{\dag}} {F \atop -G } \right), 
\end{equation} 
where $G$, $F$, and $F^{\dag}$ are the normal and anomalous Green's functions 
in the Gor'kov formalism, 
and $\zeta_{\mib p}$ is the single-particle energy measured from 
the Fermi energy. 
We divide this Green's function into two parts, 
$\widehat{g}= \widehat{g}_0+ \delta \widehat{g}$, 
where 
\begin{equation}
  \widehat{g}_0= \left( { g_0 \atop {\rm i} f_0^{\dag}} {-{\rm i}f_0 \atop -g_0 } 
  \right) 
\end{equation}
is the static part, and 
\begin{equation}
  \delta \widehat{g}= \left( { g_1 \atop -f_2} { f_1 \atop g_2 } \right)
\end{equation}
is the linear-response part. 
Then, using the Kubo formula for the charge disturbance, 
we obtain an expression for $D$ within the quasi-classical accuracy, 
\begin{equation}
  D({\rm i}\omega_m) = 
  {\rm i} 2 \pi N_{\rm F} T \sum_{\varepsilon_n} 
  \Big{\langle} 
  \widetilde{g_1} 
  \Big{\rangle}, 
\end{equation}
where $\widetilde{g_1} = {g_1}/{\Phi}$, 
and $N_{\rm F}$ is the density of states in the normal state. 
Here we neglected the contribution far from the Fermi surface.~\cite{Eliash} 
On the other hand, by a direct evaluation of the density correlation function 
in the Gor'kov formalism, we obtain, within the 
quasi-classical accuracy,~\cite{Scharnberg}   
\begin{equation}
  D({\rm i}\omega_m) =
  {\rm i} 2 \pi N_{\rm F} T \sum_{\varepsilon_n} 
  \Big{\langle} \int \frac{d \zeta_{\mib{p}}}{{\rm i} \pi} 
  ( G_0G_0-F_0F_0^\dag )
  \Big{\rangle}, 
\end{equation}
in which we introduce a shorthand notation $G_0G_0-F_0F_0^\dag = 
G_0({\rm i} \varepsilon_n+ {\rm i}\omega_m) 
G_0({\rm i} \varepsilon_n) 
- F_0({\rm i} \varepsilon_n+ {\rm i}\omega_m) 
F_0^{\dag}({\rm i} \varepsilon_n)$. 
$G_0$, $F_0$, and $F_0^{\dag}$ are the static components of the 
Green's function in the Gor'kov formalism. 
Note that, as mentioned before, we neglect the impurity vertex corrections 
in this work. 
We also neglect the vortex motion, and consider only the 
quasiparticle contribution. 
Then, by comparing the above two equations, we obtain the following relation, 
\begin{equation}
  \widetilde{g_1} = 
  \int \frac{d \zeta_{\mib{p}}}{{\rm i} \pi}  
  ( G_0G_0-F_0F_0^\dag ) .
\end{equation}

Next, we apply the same calculation scheme to the thermal conductivity $\kappa_{xx}(B)$. 
By a direct evaluation of the heat-current correlation function 
in the Gor'kov formalism, we obtain within the quasi-classical accuracy, 
\begin{eqnarray}
  Q_{xx}({\rm i}\omega_m) &=&
  -{\rm i} 2 \pi N_{\rm F} T \sum_{\varepsilon_n} 
  \Big{\langle} v_x^2 (\varepsilon_n+ \frac{1}{2}\omega_m)^2 \nonumber \\
  && \hspace{1cm}\times 
  \int \frac{d \zeta_{\mib{p}}}{{\rm i} \pi}  
  ( G_0G_0-F_0F_0^\dag ) 
  \Big{\rangle}. 
\end{eqnarray}
From the last two equations, we can express $Q_{xx}$ 
in terms of $\delta \widehat{g}$ as 
\begin{equation}
  Q_{xx}({\rm i}\omega_m) = 
  -{\rm i} 2 \pi N_{\rm F} T \sum_{\varepsilon_n} 
  \Big{\langle} v_x^2 (\varepsilon_n+ \frac{1}{2}\omega_m)^2 \; 
  \widetilde{g_1} 
  \Big{\rangle}. \label{Eq:Qxx-g1} 
\end{equation}

The quantity $\delta \widehat{g}$ is determined by the 
linear-response equation:~\cite{LO2, Ovchin} 
\begin{eqnarray}
  && \mib{v} \cdot \mib{\nabla}  \delta \widehat{g} 
  + \varepsilon_{n+} \widehat{\tau}_3  \delta \widehat{g} 
  - \varepsilon_{n}  \delta \widehat{g}  \widehat{\tau}_3 
  + {\rm i}
  \Big[|e|\mib{v}\cdot \mib{A} \widehat{\tau}_3 - \widehat{\Delta}, 
     \delta \widehat{g}  \Big]_- 
  \nonumber \\
  && 
  + \frac{1}{2 \tau}\langle \widehat{g}_0 ({\rm i} \varepsilon_{n+}) \rangle 
  \delta \widehat{g} 
  - \delta \widehat{g} \frac{1}{2 \tau} \langle \widehat{g}_0 ({\rm i}\varepsilon_{n+})
  \rangle 
  \nonumber \\
  && = 
	    \widehat{g}_0 ({\rm i} \varepsilon_{n+}) {\rm i} \Phi 
	    - {\rm i}\Phi \widehat{g}_0( {\rm i} \varepsilon_{n}), 
	    \label{Eq:linear_response}
\end{eqnarray}
where $\varepsilon_{n+}= \varepsilon_n+ \omega_m$, $[A,B]_-=AB - BA$, 
and $\widehat{\tau}_3 = ( { 1 \atop  0} {0 \atop -1 } )$. 
In the above equation, we assumed Born scattering for convenience, 
and unitarity scattering will be discussed later. 
Eq.~(\ref{Eq:linear_response}) is supplemented by the normalization condition, 
\begin{equation}
  \widehat{g}_0( {\rm i} \varepsilon_{n+}) \delta \widehat{g}+
  \delta \widehat{g} \, \widehat{g}_0( {\rm i} \varepsilon_{n}) = 0. 
  \label{Eq:normalization1} 
\end{equation}

The static part of the quasi-classical Green's function $\widehat{g}_0$ satisfies 
the following static (or Eilenberger) equation, 
\begin{equation}
  \mib{v \cdot \nabla} \widehat{g}_0 
  + \Big[ \varepsilon_n \widehat{\tau}_z +  \langle \widehat{g}_0\rangle/2 \tau 
    + {\rm i} |e| \mib{v \cdot A}- {\rm i} \widehat{\Delta}, \widehat{g}_0 \Big]_-=0, 
  \label {Eq:EE}
\end{equation}
with the normalization condition $\widehat{g}_0^2= \widehat{1}$. 
Here, 
$\widehat{\Delta}(\mib{p},\mib{r}) = 
( {0 \atop -\Delta^*_{\mib{p}}(\mib{r})} {\Delta_{\mib{p}}(\mib{r}) \atop 0} )$, 
and $\langle \cdots \rangle$ denotes the Fermi surface average. 

Once $\delta \widehat{g}$ is obtained, $\kappa_{xx}(B)$ can be calculated 
from eqs.~(\ref{Eq:Kubo}) and (\ref{Eq:Qxx-g1}). 
The Matsubara-sum in eq.~(\ref{Eq:Qxx-g1}) 
can be transformed into a real frequency integral 
using the well-known formula 
 $T \sum_{\varepsilon_n} f({\rm i} \varepsilon_n)= 
\oint \frac{d \varepsilon}{4 \pi {\rm i}}  {\rm th}(\frac{\varepsilon}{2T})  
f(\varepsilon)$. 
Then, we obtain 
\begin{equation}
  \kappa_{xx} (B)= 4 N_{\rm F} T \int_0^{\infty} 
  \frac{d \varepsilon}{2 T} 
  \frac{ \left( \frac{\varepsilon}{2 T} \right)^2 }
       { {\rm \cosh^2} \left( \frac{\varepsilon}{2T} \right) }  
       \big{\langle} v_x^2  
	       [ -{\rm i} \widetilde{g_1}^R ]_{\rm sp} 
	       \big{\rangle}, 
	       \label{Eq:Kxx}
\end{equation} 
where $\widetilde{g_1}^R(\varepsilon)= 
\widetilde{g_1}({\rm i}\varepsilon_n \to \varepsilon+ {\rm i}0_{+} )$, 
and $[\cdots]_{\rm sp}$ denotes the spatial average. 
If we perform the low-temperature expansion of the above expression, 
we have 
\begin{equation} 
  \left.\frac{\kappa_{xx}}{T} \right|_{T \to 0} 
  = \frac{\pi}{3} N_{\rm F} 
  \big{\langle} v_x^2  
      [ -{\rm i} \widetilde{g_1}^R ]_{\rm sp} 
      \big{\rangle}. 
      \label{Eq:Kxx_T0} 
\end{equation} 
Finally note that, as we will show later, the compact result~\cite{Vekhter} 
due to the BPT method 
corresponds to an approximate expression of eq.~(\ref{Eq:Kxx}).

\subsection{Numerical procedure} 
To obtain an input quantity $\widehat{g}_0$ for the linear-response 
equation, eq.~(\ref{Eq:linear_response}), we have to solve the static 
equation (\ref{Eq:EE}). 
To do so, 
we adopt the Riccati parameterization,~\cite{Scho1, Scho2}
\begin{equation}
  f_0= \frac{2a}{1+ a b}, \; 
  f^\dag_0= \frac{2b}{1+ a b}, \; 
  g_0= \frac{1- ab}{1+ a b}, 
\end{equation}
by introducing two functions $a$ and $b$. 
Then, the static equation is transformed into the following Riccati-type equations: 
\begin{eqnarray}
  \mib{v}\cdot\mib{\nabla} a 
&=&
  -2(\widetilde{\varepsilon}_n+ {\rm i}|e|\mib{v}\cdot\mib{A}) a 
  + 
  \widetilde{\Delta}_{\mib{p}}- \widetilde{\Delta}_{\mib{p}}^* a^2,  
  \label{Eq:Riccati1}\\
  \mib{v}\cdot\mib{\nabla} b 
&=&
  2(\widetilde{\varepsilon}_n+ {\rm i}|e|\mib{v}\cdot\mib{A}) b 
  - 
  \widetilde{\Delta}_{\mib{p}}^*+ \widetilde{\Delta}_{\mib{p}} b^2, 
  \label{Eq:Riccati2}
\end{eqnarray}
where $\widetilde{\varepsilon}_n= \varepsilon_n+ \langle g \rangle/2 \tau$, 
$\widetilde{\Delta}_{\mib p}= \Delta_{\mib p}+ \langle f_0 \rangle/2 \tau$, 
and $\tau$ is the quasiparticle mean free time. 
The pair potential is expressed as 
$\Delta_{\mib{p}}(\mib{r})= w_{\mib{p}} \Delta(\mib{r})$ 
with the order parameter $\Delta(\mib{r})$ and the pairing function $w_{\mib p}$. 
For a $d$-wave ($s$-wave) superconductor, we use 
$w_{\mib{p}} = \sqrt{2}(p_x^2- p_y^2)/p_{\rm F}^2$ ($w_{\mib{p}} = 1$), 
with $p_{\rm F}$ being the Fermi momentum.

The numerical procedure used to solve eqs.~(\ref{Eq:Riccati1}) and 
(\ref{Eq:Riccati2}) is described in ref.~\citen{Pedja3}. 
Calculations are performed for a two-dimensional superconductor 
with a hexagonal vortex lattice state. 
As mentioned before, we consider a moderately clean superconductor 
$\xi_0/l=0.1$, where $\xi_0 = v_{\rm F}/2 \pi T_{c0}$ and $l=v_{\rm F} \tau$, 
with $v_{\rm F}$ being the Fermi velocity. 
We initially assume a two-dimensional isotropic Fermi surface, 
and later we discuss the effect of the Fermi surface anisotropy. 
We also assume that the superconductor has a large Ginzburg-Landau parameter 
$\kappa_{\rm GL} \gg 1$ 
(most of the newly found superconductors satisfy this condition), 
so that the magnetic field $B$ is assumed to be constant. 

First, we determine the order parameter $\Delta({\mib r})$ self-consistently 
by the following gap equation, 
\begin{equation}
  \ln \left( \frac{T}{T_{c0}} \right) \Delta({\mib r}) 
  = 
  2 \pi T \sum_{\varepsilon_n>0} 
  \left( 
  \langle w_{\mib{p}} f_0 \rangle - \frac{\Delta(\mib{r})}{\varepsilon_n} 
  \right), 
\end{equation}
where $T_{c0}$ is the transition temperature at zero field without any impurity. 
This procedure requires the anomalous Green's function $f_0$, 
and it is obtained by solving 
eqs.~(\ref{Eq:Riccati1}) and (\ref{Eq:Riccati2}) on the Matsubara axis. 
Note that for a $d$-wave superconductor, we set 
the anomalous self-energy $\langle f_0 \rangle$ equal to zero. 
This treatment is consistent with the neglect of the impurity vertex corrections 
for the heat current, since both quantities vanish in the limit of 
zero magnetic field. 
Of course, for an $s$-wave superconductor, 
we cannot neglect $\langle f_0 \rangle$, 
which satisfies the Anderson theorem for pair breaking.~\cite{Anderson}  

\begin{figure}[t]
  \begin{center}
    \scalebox{0.8}[0.8]{\includegraphics{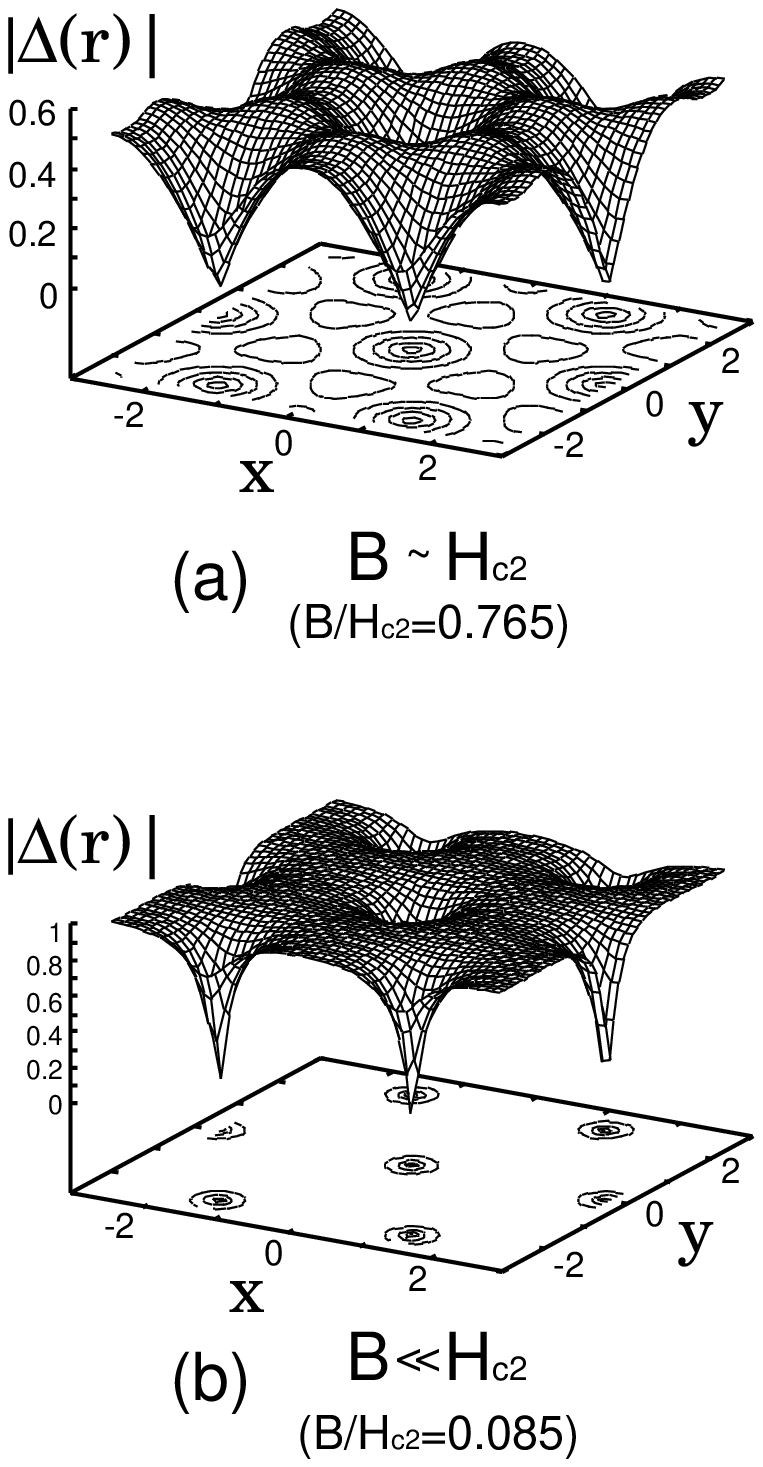}}
  \end{center}
\caption{
Real-space images of the order parameter in an $s$-wave superconductor, 
obtained from our self-consistent calculation. 
${\mib r}=(x,y)$ and $\Delta({\mib r})$  are normalized by $r_B= (2 |e| B)^{-1/2}$ 
and $\Delta_0= 1.764 T_{c0}$, respectively. 
$|\Delta({\mib r})|$ in a high field [(a)] and a low field [(b)] 
are shown. 
}
\label{Fig:Dlt-r}

\end{figure}

In Fig.~\ref{Fig:Dlt-r}, we show real-space images of the obtained order parameter 
$\Delta({\bf r})$ for an $s$-wave superconductor. 
Near $H_{c2}$ (Fig.~\ref{Fig:Dlt-r}(a)), vortices overlap with each other. 
However, upon lowering the field $H \ll H_{c2}$ (Fig.~\ref{Fig:Dlt-r}(b)), 
the vortex core radius becomes smaller and smaller 
than the intervortex spacing. 
These behaviors can be obtained only after we 
calculate the order parameter self-consistently. 
After we determine the order parameter $\Delta({\mib r})$, we solve 
eqs.~(\ref{Eq:Riccati1}) and (\ref{Eq:Riccati2}) 
for the real frequency ${\rm i}\varepsilon_n \to \varepsilon \pm {\rm i}0_+$ 
to obtain the retarded (advanced) Green's function. 
The impurity self-energy is determined self-consistently 
in this static equation.~\cite{Pedja3} 

Next we solve the linear-response equation. 
To calculate $\kappa_{xx}(B)$ using eq.~(\ref{Eq:Kxx}), 
we need to carry out the analytical continuation of eq.~(\ref{Eq:linear_response}) 
to real frequencies. 
The off-diagonal components of $\delta \widehat{g}$ can be eliminated 
by the normalization condition, eq.~(\ref{Eq:normalization1}). 
Then, the diagonal components of the linear-response equation, 
eq.~(\ref{Eq:linear_response}), yield 
\begin{equation}
  ( \widehat{L}+ \widehat{M} ) 
\left(  
\begin{array}{l} 
  g_1\\
  g_2\\
\end{array}  
\right) = 
{\rm i}\Phi(g^R- g^A) 
\left(  
\begin{array}{l}
  1\\
  1\\
\end{array} 
\right)
.   \label{Eq:g1g2_v01} 
\end{equation}
Here, $\widehat{L}$ and $\widehat{M}$ are defined by 
\begin{eqnarray}
  \widehat{L} &=& 
  \left( 
  \begin{array}{cc}
    \mib{v} \cdot \mib{\nabla} & 0 \\ 
    0 & - \mib{v} \cdot \mib{\nabla}\\ 
  \end{array} \right),  \\
  \widehat{M} &=& 
  \left( 
  \begin{array}{cc}
    \sigma+ \alpha  & \beta \\
    \alpha  & \sigma+ \beta \\
  \end{array} \right), 
\end{eqnarray}
where 
$\alpha(\mib{r})= \frac{ \widetilde{\Delta}_{\mib p} f_0^{\dag R}
  + \widetilde{\Delta}_{\mib p}^* f_0^A }
{g_0^R- g_0^A }$, 
$\beta(\mib{r})= \frac{ \widetilde{\Delta}_{\mib p} f_0^{\dag A}
  + \widetilde{\Delta}_{\mib p}^* f_0^R }
{g_0^R- g_0^A } $, 
$\sigma(\mib{r})= \frac{1}{2 \tau} 
\langle g_0^R- g_0^A \rangle$, 
and $\widetilde{\Delta}_{\mib p}= \Delta_{\mib p}+ 
\frac{1}{2 \tau} \langle f_0^R \rangle $. 

To solve eq.~(\ref{Eq:g1g2_v01}), we first divide $\widehat{M}$ into two parts, 
$\widehat{M}= [\widehat{M}]_{\rm sp}+\delta{\widehat{M}}$, and 
rewrite eq.~(\ref{Eq:g1g2_v01}) as 
\begin{equation}
  (\widehat{L}+ [\widehat{M}]_{\rm sp} ) 
  \left(  
  \begin{array}{l}
    g_1\\
    g_2\\
  \end{array} 
  \right) 
  =
  {\rm i}\Phi(g^R- g^A) 
  \left(  
  \begin{array}{l}
    1\\
    1\\
  \end{array} 
  \right) 
  - 
  \delta \widehat{M} 
    \left(  
  \begin{array}{l}
    g_1\\
    g_2\\
  \end{array} 
  \right). 
  \label{Eq:g1g2_v02}
\end{equation}
Since $g_1$ and $g_2$ are gauge-invariant and periodic, 
eq.~(\ref{Eq:g1g2_v02}) can be solved iteratively 
by a Fourier transform.~\cite{Pedja4} 

\subsection{Relation to the BPT approximation} 
For an analytical treatment, it is convenient to introduce the 
BPT approximation~\cite{BPT,Pesch}, 
which has been applied to the approximate calculation of the thermal 
conductivity.~\cite{Houghton,Vekhter} 
Recently, this BPT approximation has been used frequently 
to interpret experimental data.~\cite{Tewordt,Kusunose} 
To derive the compact result using the BPT approximation, 
we first neglect the spatial variations of $g_1$ and $g_2$ in eq.~(\ref{Eq:g1g2_v01}) 
and remove $\widehat{L}$. 
Next, we perform a spatial average on each matrix element of $\widehat{M}$ 
as $\alpha(\mib{r}) \to 
\frac{ [\widetilde{\Delta}_{\mib p} f_0^{\dag R}
    + \widetilde{\Delta}_{\mib p}^* f_0^A ]_{\rm sp}}
{[g_0^R- g_0^A]_{\rm sp} }$, 
$\beta(\mib{r}) \to  
\frac{[ \widetilde{\Delta}_{\mib p} f_0^{\dag A}
    + \widetilde{\Delta}_{\mib p}^* f_0^R ]_{\rm sp}} 
{[g_0^R- g_0^A ]_{\rm sp}} $, 
and 
$\sigma( \mib{r}) \to 
\frac{1}{2 \tau} 
[\langle g_0^R- g_0^A \rangle]_{\rm sp}$. 
Note that the spatial average is performed separately 
in the denominator and the numerator. 
Then, $g_1$ and $g_2$ can be obtained by a matrix inversion, and this yields 
\begin{eqnarray}
  \kappa_{xx}(B) 
  &=&
  8 N_{\rm F} T 
  \int_0^\infty 
  \frac{d \varepsilon}{2 T} 
  \frac{ \left( \frac{\varepsilon}{2 T} \right)^2 } 
       { {\rm \cosh^2} \left( \frac{\varepsilon}{2T} \right) }  \nonumber \\
       && \qquad \times
       \big{\langle} 
       v_x^2 {\rm Re}[g_0^R]_{\rm sp} 
       \tau^{\rm tr}_{\mib p} (\varepsilon) \big{\rangle}, \label{Eq:BPT1} \\ 
       \frac{1} {\tau^{\rm tr}_{\mib{p}} (\varepsilon) }
  &=& 
       \frac{1}{\tau} {\rm Re}[g^R]_{\rm sp} 
  + 
  \frac{{\rm Re}[\widetilde{\Delta}_{\mib{p}}f_0^{\dag R}
      +\widetilde{\Delta}_{\mib{p}}^* f_0^{R} ]_{\rm sp}}
	{ {\rm Re}[g_0^R]_{\rm sp} } . 
	\label{Eq:BPT2}
\end{eqnarray} 
If we substitute the result of the BPT approximation for the static 
components of the quasi-classical Green's function, 
then the above expression reproduces the compact result derived 
in ref.~\citen{Vekhter}. 
 
\section{Numerical Results} 

\begin{figure}[t]
  \scalebox{0.55}[0.55]{\includegraphics{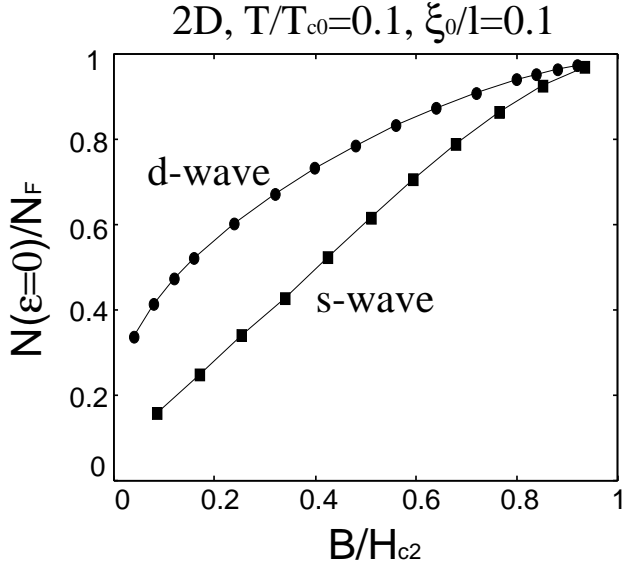}}
\caption{Magnetic-field dependence of zero-energy density of states $N(\varepsilon=0)$ 
in $s$-wave (filled squares) and $d$-wave (filled circles) superconductors 
in the moderately clean case ($\xi_0/l=0.1$). 
Here, $N_{\rm F}$ is the density of states in the normal state.} 
\label{Fig:Gav-h0}
\end{figure}

First, we show our numerical results for the density of states. 
The density of states in a superconductor is given by 
\begin{equation}
  \frac{N(\varepsilon)}{N_{\rm F}}
  =
  {\rm Re}[ \langle g_0^R(\varepsilon) \rangle ]_{\rm sp}. 
\end{equation}
Of particular interest is the zero-energy density of states 
$N(\varepsilon=0)$, i.e., the density of states at the Fermi energy. 
This is because $N(\varepsilon=0)$ is related to the low-temperature specific 
heat $C$ through 
${N(\varepsilon=0)}/{N_{\rm F}} = C/C_N |_{T \to 0}$, 
where $C_N$ is the normal-state specific heat. 

In Fig.~\ref{Fig:Gav-h0}, 
we plot the zero-energy density of states in 
$s$-wave and $d$-wave superconductors ($\xi_0/l=0.1$) 
versus magnetic field $B$. 
As in the impurity-free case,~\cite{Nakai} 
we can see the well-known difference in their field dependences 
even in the moderately clean superconductor. 
For an $s$-wave superconductor, we find a linear dependence of 
$N(\varepsilon=0)/N_{\rm F}$ on $B$, which originates from 
the localized quasiparticles within the vortex cores, 
while for a $d$-wave superconductor, we find 
an approximate $\sqrt{B}$ dependence due to the 
field-induced Doppler shift on the delocalized quasiparticles. 
A feature specific to the impure case is that 
there is a residual value of $N(\varepsilon=0)$ in the zero-field ($B \to 0$) limit. 

\begin{figure}[t]
  \scalebox{0.55}[0.55]{\includegraphics{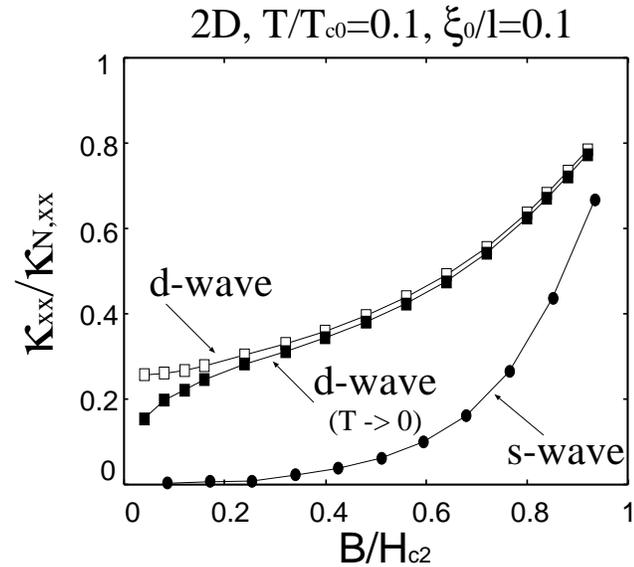}}
\caption{
  Magnetic-field dependences of the longitudinal thermal conductivity 
  $\kappa_{xx}(B)$ in $s$-wave (filled circles) and $d$-wave (open squares) 
  superconductors. 
  $\kappa_{{\rm N},xx}=\frac{1}{3}\pi^2v_{\rm F}^2 T \tau$ is the 
  thermal conductivity in the normal state. 
  Filled squares represent the zero-temperature value of $\kappa_{xx}(B)$ 
  [see eq.~(\ref{Eq:Kxx_T0})] in a $d$-wave superconductor. 
}
\label{Fig:Kij-h0}
\end{figure}

Next, we show our numerical result for the thermal conductivity $\kappa_{xx}(B)$. 
Figure \ref{Fig:Kij-h0} shows the magnetic-field dependences of 
the calculated thermal conductivities in $s$-wave and $d$-wave superconductors. 
In the figure, we can observe a considerable difference in their field dependences: 
in an $s$-wave superconductor, $\kappa_{xx}(B)$ at low fields 
is exponentially small, 
while in a $d$-wave superconductor, there is a finite amount of thermal transport 
even at low fields. 
This means that the field dependence of $\kappa_{xx}(B)$ can be used 
as a method of probing gap nodes. 
However, we would like to emphasize here that unless we discuss the 
zero-temperature limit, $\kappa_{xx}(B)$ in both superconductors have similar  
field dependences. 
This is evident when we look at the data in Fig.~\ref{Fig:Kij-h0} 
at finite temperatures. 
Actually, it is difficult to distinguish $\kappa_{xx}(B)$ in a $d$-wave superconductor 
from that in an $s$-wave superconductor except for their residual values, 
although the data are plotted in a moderately low temperature region ($T/T_{c0}=0.1$). 
Therefore, if we aim to use $\kappa_{xx}(B)$ to detect gap nodes, 
we have to extract its value at zero temperature. 

\begin{figure}[t]
  \scalebox{0.55}[0.55]{\includegraphics{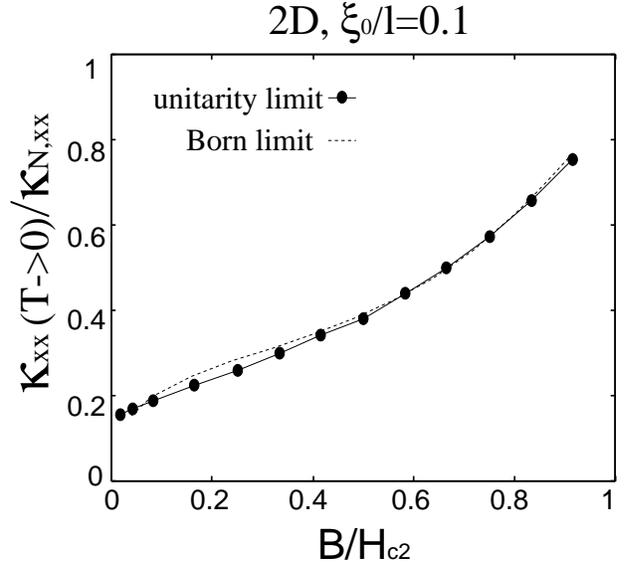}} 
\caption{
  Magnetic-field dependence of zero-temperature thermal conductivity 
  in a $d$-wave superconductor in the unitarity limit. 
}
\label{Fig:Kij-h0_Ulimit}
\end{figure}

In some unconventional superconductors, 
thermal conductivity measurements in zero magnetic field 
suggest the necessity of including the resonant impurity scattering 
in the unitarity limit.~\cite{Pethick,Hirschfeld2,Schimitt} 
This effect can be taken into account 
by performing the replacement 
$\frac{1}{2 \tau} \langle g_0 \rangle 
\to \frac{1}{2 \tau} \langle g_0 \rangle^{-1}$ 
in eqs.~(\ref{Eq:linear_response}) and (\ref{Eq:EE}). 
Figure~\ref{Fig:Kij-h0_Ulimit} shows the magnetic-field dependence of 
zero-temperature thermal conductivity in a $d$-wave superconductor 
with the unitarity-limit scattering. 
The result suggests that 
the magnetic-field dependence of $\kappa_{xx}(B)$ 
in the unitarity limit is slightly different from that in the Born limit, 
particularly at low fields, 
while the main conclusion of the previous paragraph remains unchanged.

Here, we comment on the difference between the present result and 
that of the BPT approximation. 
In the BPT approximation, the spatial average is performed incoherently 
as noted above using eq.~(\ref{Eq:BPT1}). 
This procedure is expected to overestimate the vortex-core scattering 
and reduce the thermal conductivity. 
Actually, if we compare Fig.~1 of ref.~\citen{Vekhter} with the present result, 
we can see that $\kappa_{xx}(B)$ obtained by the BPT approximation 
is slightly suppressed to a lower value. 
It has weak field dependences in the intermediate fields, 
and shows a steep increase near $H_{c2}$. 
In the present study, in contrast, 
we do not use such an artificial spatial averaging procedure, 
and we take account of the spatial dependence of Green's function 
in both static and linear-response equations.

A careful reader may think that 
the inclusion of Fermi surface anisotropy may strongly affect 
the result. 
Indeed, we have realized that an in-plane Fermi surface anisotropy 
{\it does} affect the field-angle-dependent specific heat 
and magnetization oscillations.~\cite{Pedja1b,Adachia} 
To consider this possibility, 
we investigate the effect of in-plane Fermi surface anisotropy on 
the magnetic-field dependence of $\kappa_{xx}(B)$.  
Let us consider a model dispersion,
\begin{equation}
  \epsilon_{\mib{p}}= \frac{1}{2m}(p_x^2+ p_y^2)\big( 1+ b \cos(4 \varphi) \big), 
  \label{Eq:b-model}
\end{equation}
where $\varphi$ denotes the azimuthal angle of Fermi momentum $\mib{p}$, 
and the parameter $b$ represents the 
degree of in-plane Fermi surface anisotropy with a fourfold symmetry. 
Here, we consider a case where the superconducting gap is an isotropic $s$-wave-type 
but the Fermi surface has in-plane anisotropy. 

\begin{figure}[t]
  \scalebox{0.55}[0.55]{\includegraphics{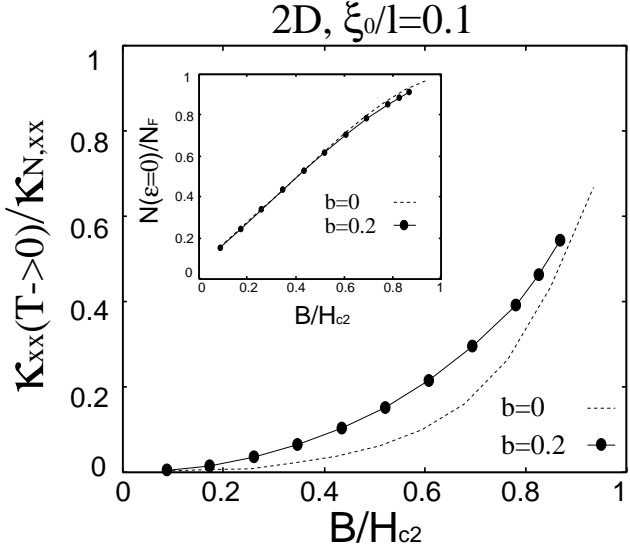}}
\caption{
  Magnetic-field dependences of $\kappa_{xx}$ 
  in an $s$-wave superconductor with an anisotropic Fermi surface. 
  The inset shows the corresponding data for $N(\varepsilon=0)/N_{\rm F}$. 
  The definition of the anisotropy parameter $b$ is given in eq.~(\ref{Eq:b-model}). 
}
\label{Fig:GavKij_Bmodel}
\end{figure}

Figure~\ref{Fig:GavKij_Bmodel} shows the magnetic-field dependences of 
$\kappa_{xx}(B)$ for $b=0.2$. 
Note that the degree of anisotropy ($b=0.2$) 
is sufficiently strong to reverse the in-plane magnetization oscillation 
pattern~\cite{Adachia} in a $d$-wave superconductor. 
Concerning the field dependence of $\kappa_{xx}(B)$, there is a 
difference between results for $b=0$ and $b=0.2$. 
However, the difference is quantitative, whereas the 
difference between $s$-wave and $d$-wave results (Fig.~\ref{Fig:Kij-h0}) 
is qualitative. 
This means that the superconducting gap structure 
influences the field dependence of $\kappa_{xx}(B)$ more strongly 
than the Fermi surface anisotropy. 

\section{Discussion} 
\begin{figure}[t]
  \scalebox{0.55}[0.55]{\includegraphics{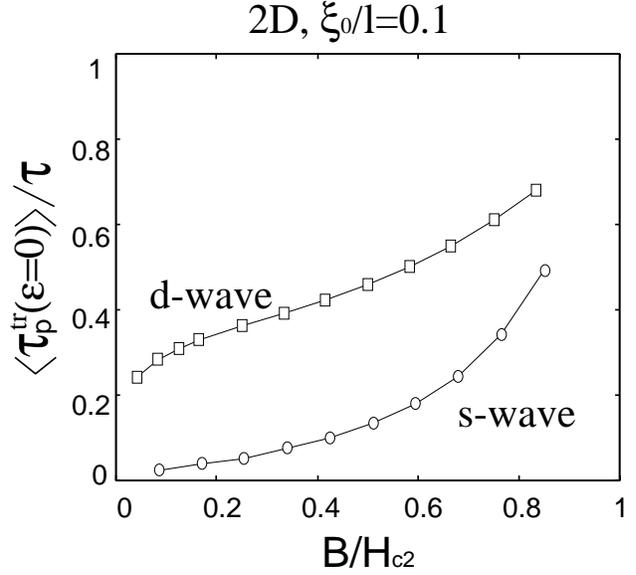}}
\caption{
  Magnetic-field dependence of $\langle \tau^{\rm tr}_{\mib p}(\varepsilon=0) \rangle$ 
  in $s$-wave and $d$-wave superconductors. 
}
\label{Fig:Tau-h}
\end{figure}

We discuss the origin of the difference in $\kappa_{xx}(B)$ 
between a $d$-wave superconductor and an $s$-wave superconductor; 
a $d$-wave superconductor can transport a certain amount of heat even 
at low fields, whereas 
in an $s$-wave superconductor, the thermal transport 
is completely suppressed. 
For this purpose, it is convenient to 
employ the zero-temperature expression of eq.~(\ref{Eq:BPT2}), 
\begin{equation}
  \left. \frac{\kappa_{xx}}{T} \right|_{T \to 0} 
  =
  \frac{2}{3} N_F
  \big{\langle} 
  v_x^2 {\rm Re}[g_0^R]_{\rm sp} 
  \tau^{\rm tr}_{\mib p} (\varepsilon=0) \big{\rangle}. 
  \label{Eq:Boltzmann}
\end{equation}
In Fig.~\ref{Fig:Tau-h}, we show the magnetic-field dependence of 
$\langle \tau^{\rm tr}_{\mib{p}}(\varepsilon =0) \rangle$ in both superconductors. 
By comparing this figure with Fig.~\ref{Fig:Kij-h0}, we can conclude 
that the main difference in $\kappa_{xx}(B)$ between $s$-wave 
and $d$-wave superconductors results from that in the transport lifetime 
$\langle \tau^{\rm tr}_{\mib p}(\varepsilon=0) \rangle$.

We can explain this difference in the following way. 
In an $s$-wave superconductor, the Andreev scattering rate 
given by the second term on the right-hand side of eq.~(\ref{Eq:BPT2}) 
increases rapidly upon 
decreasing the field due to the fact that it contains the small factor 
${\rm Re}[g_0^R]_{\rm sp} \sim B/H_{c2}$ in the denominator. 
In a $d$-wave superconductor, quasiparticles moving along the antinodal direction 
are dominated by the same principle. 
However, nodal quasiparticles do not undergo the Andreev scattering mechanism 
since ${\rm Re}[\widetilde{\Delta}_{\mib p} f_0^{\dag R} + 
  \widetilde{\Delta}^*_{\mib p} f_0^{R}]_{\rm sp}$ is almost zero, 
and they can transport a certain amount of heat. 
It is this effect that yields the difference in $\kappa_{xx}(B)$ 
between a nodal superconductor and a fully gapped superconductor. 

This interpretation is consistent with the following physical picture 
provided by Boaknin {\it et al.}~\cite{Boaknin1} 
In an $s$-wave superconductor, localized quasiparticles within vortices, 
which are relevant to the low-temperature properties, 
can give a linear-field density of states, 
but cannot contribute to the heat transport due to their localized nature. 
In a $d$-wave superconductor, on the other hand, 
Doppler shifted nodal quasiparticles can contribute to the heat transport, 
as well as to the well-known $\sqrt{B}$-dependent density of states. 
Thus, $\kappa_{xx}(B)$ can pick up only the contribution of 
extended quasiparticles. 
To some extent, this picture can be inferred from 
``extended zero-energy density of states $N_{\rm ext}(B)$'' 
defined in ref.~\citen{Nakai} (see Fig.~1 therein). 
Our study confirmed this physical picture by a concrete numerical calculation. 

In this work, we only consider a single-band superconductor 
with a two-dimensional Fermi surface. 
Hence, we do not present a detailed comparison with experiments. 
Instead, we discuss the implications of the present results on experiments. 
For $s$-wave superconductors, the present result is able to explain 
the following experimental observations. 
The magnetic-field dependence of $\kappa_{xx}(B)$ in 
Fig.~\ref{Fig:Kij-h0} has a good correspondence to 
the experimental data for Nb (Fig.~3 of ref.~\citen{Lowell}) and V$_3$Si 
(Fig.~3(c) of ref.~\citen{Boaknin2}), although our 
result cannot be applied directly to the low-$\kappa_{\rm GL}$ 
superconductor Nb. 
Recently, Kasahara {\it et al.} also observed a similar magnetic-field 
dependence of $\kappa_{xx}(B)$ in a pyrochlore superconductor 
KOs$_2$O$_6$.~\cite{Kasahara} 
Concerning the $d$-wave calculation, 
our result is consistent with the thermal conductivity of 
Bi$_2$Sr$_2$CaCu$_2$O$_8$ at low fields.~\cite{Aubin} 
If we limit ourselves to the low-field region, 
our result is also consistent with the 
experimental data in other (perhaps nodal) 
superconductors,~\cite{Suderow,Boaknin1,Izawa} 
in that these data at low fields have stronger field dependences 
than the $s$-wave superconductors discussed above. 
On approaching the upper critical field, however, 
a marked difference appears. 
Our result shows $(H_{c2}-B)^{1/2}$ behavior near $H_{c2}$, 
which is consistent with the previous work,~\cite{Maki2} 
while the experimental data show $(H_{c2}-B)^\eta$ dependences with 
$\eta \ge 1$. 
We leave this discrepancy to a future work. 

\section{Conclusion} 
In this paper, we have studied the magnetic-field dependence of 
the mixed-state thermal conductivity $\kappa_{xx}(B)$ 
in $s$-wave and $d$-wave superconductors 
based on the quasi-classical theory of superconductivity. 
The present work is valid in a wide field range from $H_{c1}$ to $H_{c2}$, 
and beyond the previous theoretical studies; 
the Doppler shift approximation is valid only near $H_{c1}$, 
and the Brandt-Pesch-Tewordt approximation is valid only near $H_{c2}$. 

On the basis of our result, we clarified that 
there is indeed a clear difference 
in the low-temperature mixed-state thermal transport 
between $s$-wave and $d$-wave superconductors; 
in a $d$-wave superconductor, there is a finite amount of thermal transport 
even at low fields, 
while in an $s$-wave superconductor, the thermal transport is completely suppressed. 
This suggests that the magnetic-field dependence of $\kappa_{xx}(B)$ can be 
used to distinguish a nodal superconductor from a fully gapped superconductor, 
on the condition that we extract a suitable zero temperature-limit. 
Also, we microscopically confirmed the physical picture provided 
by Boaknin {\it et al.}~\cite{Boaknin1} 
on the mixed-state thermal transport, 
and demonstrated that the above-mentioned difference in $\kappa_{xx}(B)$ 
mainly results from the corresponding difference in the transport lifetime.

\acknowledgements
One of the authors (H. A.) would like to thank H. Ebisawa, 
Y. Matsuda, T. Shibauchi, and Y. Kasahara for discussions, 
and M. Eschrig for important comments. 
He is also grateful to Y. Nisikawa and A. Furukawa 
for introducing him to the application of Kubo formalism 
to thermal transport. 
A part of the numerical calculations was carried out 
on Altix3700 BX2 at YITP in Kyoto University.


\begin{thebibliography}{99}
\bibitem{comm1}
  In the Meissner state ($B=0$), a Ward identity ensures 
  that the impurity vertex corrections vanish for 
  even-parity superconductors.~\cite{Ambegaokar,Hirschfeld}
\end{thebibliography}

\end{document}